%% file: rm.tex
\documentclass[11pt]{article}

\pdfoutput=1

\usepackage{amsmath}
\usepackage{amssymb}
\usepackage{mathtools}
\usepackage{amsfonts}
\usepackage{graphicx}
\usepackage{subcaption}
\newcommand{\edge}{\mathfrak{e}}

\begin{document}

\title{Energy landscapes for the self-assembly of supramolecular polyhedra\thanks{ERR acknowledges the support and hospitality of the Institute for Computational and Experimental Research in Mathematics (ICERM) at Brown University, where she was a postdoctoral fellow while carrying out this work.
 GM acknowledges partial support from NSF grant DMS 14-11278. This research was conducted using computational resources at the Center for Computation and Visualization, Brown University.}}

\author{Emily R. Russell\thanks{ICERM, Box 1955, Brown University, Providence, RI 02912, USA ({\tt emily\_russell@post.harvard.edu}).}, Govind Menon\thanks{Division of Applied Mathematics, Brown University, 182 George St., Providence, RI 02912, USA ({\tt govind\textunderscore menon@brown.edu}).}}

\maketitle


\begin{abstract} 
We develop a mathematical model for the energy landscape of polyhedral  supramolecular cages recently synthesized by self-assembly~\cite{Fujita1}. Our model includes two essential features of the experiment: (i) geometry of the organic ligands and metallic ions; and (ii) combinatorics. The molecular geometry is used to introduce an
energy that favors square-planar vertices (modeling  $\mathrm{Pd}^{2+}$ ions) and bent edges with one of two preferred opening angles (modeling boomerang-shaped  ligands of two types).   The combinatorics of the model involve $2$-colorings of edges of polyhedra with $4$-valent vertices. The set of such $2$-colorings, quotiented by the octahedral symmetry group, has a natural graph structure, and is called the combinatorial configuration space. The energy landscape of our model is the energy of each state in the 
combinatorial configuration space.

The challenge in the computation of the energy landscape is a combinatorial explosion in the number of $2$-colorings of edges. We describe sampling methods based on the symmetries of the configurations and connectivity of the configuration graph. When the two preferred opening angles encompass the geometrically ideal angle, the energy landscape exhibits a very low-energy minimum for the most symmetric configuration at equal mixing of the two angles, even when the average opening angle does not match the ideal angle.

\end{abstract}




\input{intro}

\input{config}

\input{sampling}
\input{results}

\input{conclusions}

\bibliographystyle{siam}
\bibliography{rm}
\end{document}

%% file: intro
\section{Introduction}
\label{sec:intro}
The central goal of the field of synthetic self-assembly is to create ordered assemblies of building blocks by mimicking the essentials of biophysical processes of self-assembly. In supramolecular chemistry, the building blocks are molecules, and a basic problem is to synthesize supramolecular cages, to shed light on the self-assembly of viral capsids~\cite{Ward,Fujita,CK1}.  Such cages may be used to contain or catalyze chemical reactions with great precision. For example,  molecular flasks that stabilize highly reactive compounds,  such as phosphorus, have been synthesized by self-assembly~\cite{trap-fire}. While the chemistry of these examples differ, all of them share fundamental geometrical features. The cages are  Platonic or Archimedean polyhedra, and their molecular design  relies on the decomposition of the polyhedra into simpler geometric units, and a careful choice of molecules whose reactive units can be idealized as such units. For example, in the experiments reported in~\cite{Ward}, the fundamental units are hexagonal `molecular tiles' that assemble into (quasi)-truncated octahedra,  each molecular tile forming one hexagonal face of the truncated octahedron. 

Our primary interest in this article is to develop a mathematical model that sheds light on the self-assembly of organometallic molecular cages synthesized in  Fujita's lab~\cite{Fujita1}. These cages may be idealized as Platonic or Archimedean solids with $4$-valent vertices. They are realized in experiments by organometallic complexes consisting of $4$-valent palladium ($\mathrm{Pd}^{2+}$) ions at vertices linked by `boomerang-shaped' organic ligands. All such molecular cages have the general chemical formula $M_{n}L_{2n}$. The subscript of $M$ indicates the number of palladium ions at the vertices, and the subscript of $L$ denotes the number of organic ligand molecules making up the edges of the polyhedron. The symmetry of the vertices forces $n=6$, $12$, $24$, $30$ and $60$.  Several of these ligand polyhedra were observed in experiment, including the octahedron ($n=6$), cuboctahedron ($n=12$), and rhombicuboctahedron ($n=24$)~\cite{Fujita1}. 

A particularly interesting effect was observed when two different ligand molecules  with differing bend angle  were mixed.  The concentration of the two different organic ligand molecules, types $A$ and $B$ say, served as a control parameter.  Separately, these ligands formed cuboctahedra and  rhombicuboctahedra  of the form $M_{12}A_{24}$ and $M_{24}B_{48}$ respectively. In a mixture, a variety of supermolecules with the general chemical formula  $M_n A_{2n-m} B_{m}$, $0 \leq m \leq 2n$, and $n=12$ or $n=24$ are theoretically possible. Further, for $1 \leq m \leq 2n-1$ these polyhedral supermolecules typically have many distinct isomers, corresponding to different geometric arrangements of the $A$ and $B$ ligands. In experiment, as the concentration ratio was varied, a sharp transition was observed from a solution in which only cuboctahedra formed to one in which only rhombicuboctahedra formed. More precisely,  while a solution consisting of all isomers $M_{n}A_{2n-m}B_{m}$ with $n=12$ or $n=24$ and $0 \leq m \leq 2n$ is possible in theory, in each experiment with a fixed concentration ratio of $A$ and $B$ ligands, the equilibrium solution consisted of only cuboctahedral isomers ($n=12$) or only rhombicuboctahedral isomers ($n=24$).  The sharpness of this phase transition was described as {\em emergent behavior\/} by Fujita and co-authors~\cite{Fujita1}. In order to  understand what drives this phase transition, we develop a `minimal' model that combines the geometry of  isomers with a phenomenological energy for deviations from the ideal bend angle. We then compute an energy landscape for isomers with the general chemical formula $M_{n}A_{2n-m}B_{m}$, $n=6$, $12$ and $24$, $0 \leq m \leq 2n$.

Despite the apparent specificity of our model, the ideas  presented here may be naturally adapted to the other experiments. More broadly, the development of a theoretical understanding of  self-assembly is a deep fundamental exercise, rich in biological, mathematical and physical ideas. Our work is an instance of small, but growing, mathematical literature on synthetic  self-assembly that emphasizes ideas from discrete geometry and statistical physics~\cite{CH,MG1,MG2}. These models typically include three aspects. The first is the identification of a configuration space that idealizes the set of intermediate structures that lie between the building blocks and the assembled product. The second is to understand the free energy landscape; in particular, to understand which of the states in the configuration space  are energetically favorable. The third is a description of the kinetics of self-assembly. In this article, we focus on the first two aspects of this procedure, since (as will become clear) determining the energy landscape  requires intensive computation, and is of independent interest. 

The rest of this article is organized as follows. We describe the configuration space and energy in the next section. This is followed by a description of the combinatorial explosion, and a symmetry-based sampling scheme. Finally, we describe our numerical results and conclusions.

%% file: config
\section{The configuration space and energy function}
\label{sec:config-space}
We are interested in molecules of the form $M_{n}A_{2n-m}B_{m}$, with $n$ vertices corresponding to the ions $M$, $2n-m$ ligands of type $A$, and $m$ ligands of type $B$. Our model involves: (i) simple geometric combinatorics, (ii) a phenomenological energy function, and (iii) sampling schemes for fast computation.
(i) For any $m$, the number of distinct isomers is the number of distinct $2$-colorings of the edges of the polyhedra modulo the action of  the symmetry group of the polyhedron. (ii) Our energy function  penalizes the geometric distortion of the polyhedra caused by ligand molecules with differing bend angles. (iii) A calculation of the full energy landscape is impossible for $n=12$ and $n=24$ because of a combinatorial explosion. We use symmetry and heuristics based on $n=6$ to sample the energy landscape.

\subsection{Colorings and configurations}
\begin{figure}
\centering
\includegraphics[width=\textwidth,natwidth=1095,natheight=1095]{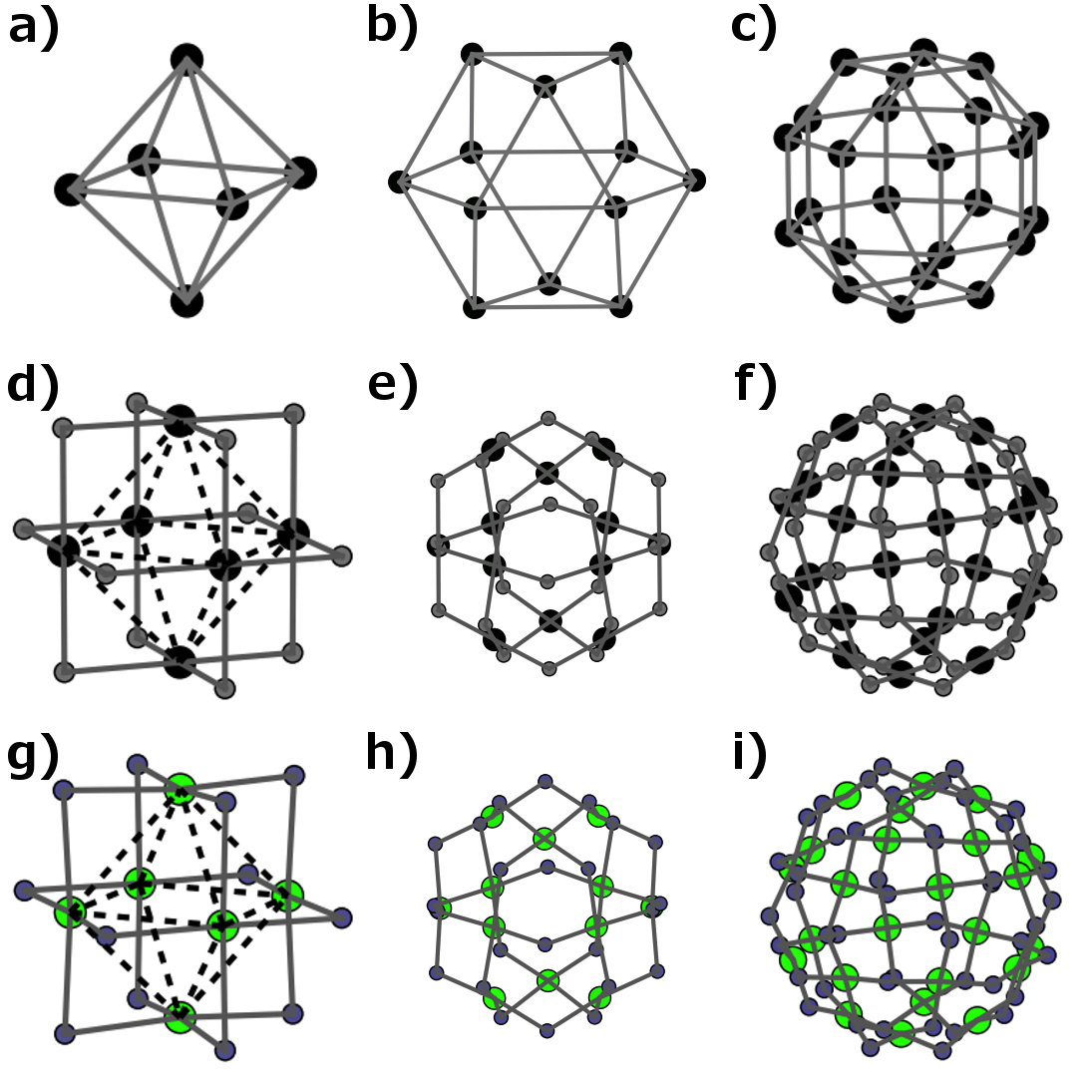}
\caption{{\bf $4$-valent polyhedra with $O_h$ symmetry.\/} a) octahedron; b) cuboctahedron; c) rhombicuboctahedron.   (d-f) {\bf Ideal supramolecular cage embeddings.\/}  Here, the $\mathrm{Pd}^{2+}$ ions are represented by large black vertices, while the `boomerang'-shaped organic ligands are represented by small gray vertices (`elbows'), each with two arms; in (d), the straight edges of the octahedron are included only for comparison.  Note that the geometry at the ions is square-planar.  (g-i) {\bf Frustration.\/} Minimum-energy embeddings when the preferred bend angle of the edges is $\theta_0 < \theta_*$.  g) $\theta_0 = 81^\circ$, $\theta_* = 90^\circ$; h) $\theta_0 \approx 105.5^\circ$, $\theta_* \approx 117.2^\circ$; i) $\theta_0 \approx 120.7^\circ$, $\theta_* \approx 134.1^\circ$.  Brightness of color indicates the energy contribution from each vertex and each `elbow'.  Note that the geometry at the ions is deformed from square-planar.}
\label{Fig:polyhedra}
\end{figure}

More formally, let $\mathcal{P}$ be a polyhedron whose faces, edges and vertices are denoted $(\mathcal{F},\mathcal{E},\mathcal{V})$. We assume that $\mathcal{P}$ is
one of the following: the octahedron; cuboctahedron; or rhombicuboctahedron. All these polyhedra, shown in Fig.~\ref{Fig:polyhedra}, have regular $4$-valent vertices and their symmetry group is the octahedral group, $O_h$~\cite{Coxeter}. A $2$-coloring of edges, or coloring for short, is a map $c: \mathcal{E} \to \{0,1\}$. The set of colorings is denoted $C$. It is convenient to view the colorings `physically': we identify the polyhedron $\mathcal{P}$ with its standard embedding in space, and paint  the edges of $\mathcal{P}$ blue ($0$) or red ($1$), for ligands of type $A$ and $B$ respectively. We say that two colorings $c_1$ and $c_2$ are equivalent under $O_h$, written $c_1 \sim c_2$, if there exists $g \in O_h$ such that $gc_1=c_2$. We define a {\em configuration\/} to be an equivalence class of colorings under the relation $\sim$. The quotient set $C/O_h$ is called the {\em combinatorial configuration space\/}, or simply {\em configuration space\/}. It is the set of all distinct colorings modulo the symmetry of the polyhedron.

The enumeration of $C/O_h$ is a classic problem in discrete group theory (we  follow~\cite{Rotman}). Let $O_h c$ denote the orbit of a coloring $c$. We say that the {\em degeneracy\/} of a coloring $c$ is the size of the orbit $|O_h c|$. Clearly, the degeneracy is the same for all colorings in the equivalence class $[c]$. Thus, it is meaningful to speak of  the degeneracy of $[c]$. Let $o_c \subset O_h$ denote the stabilizer subgroup of $c \in C$. We call the size of this subgroup, $|o_c|$,  the {\em symmetry number\/} of $c$, and denote it by $s$.  By the orbit-stabilizer theorem and Lagrange's theorem, for every coloring $c \in C$,
\begin{equation}
\label{eq:Lagrange}
|O_h c| \times |o_c| =  |O_h| = 48.
\end{equation} 

Two colorings $c_1$ and $c_2$ are equivalent if and  only if they lie on the same group orbit. Thus, the size of the configuration space, $|C/O_h|$, is simply the number of distinct orbits. Let $C^g$ denote the set of colorings fixed by $g \in O_h$, i.e. $C^g = \{c \in C\left| gc=c \right.\}$.  By Burnside's lemma,
\begin{equation}
\label{eq:burnside}
|C/O_h| = \frac{1}{|O_h|} \sum_{g \in O_h} |C^g|. 	
\end{equation}

While this  calculation of the size $|C/O_h|$ is well-known, our interest lies in an explicit description of each equivalence class $[c] \in C/O_h$. We compute $C/O_h$ explicitly for the octahedron and cuboctahedron as described below.  We are unable to explicitly compute $C/O_h$ for the rhombicuboctahedron because of a combinatorial explosion. Note that $|C/O_h|$ is bounded below by
\begin{equation}
\label{eq:estimate}	
|C/O_h| \geq \frac{|C|}{|O_h|} = \frac{2^{|\mathcal{E}|}}{48}.
\end{equation}
The octahedron has 144 unique configurations  (the lower bound is $85$) and  the cuboctahedron has 352,744 unique configurations  (the lower bound is $349,525)$. For the rhombicuboctahedron, the same bound yields $2^{48}/48 \approx 5,864,062,014,806$, which is  too large to enumerate explicitly. Thus, one of the main computational challenges in our work is to obtain a realistic understanding of the energy landscape, despite the combinatorial explosion of $|C/O_h|$.

Finally, note that the configuration space $C/O_h$ inherits a natural graph structure from $C$. We say that two colorings $c_m$ and $c_{m+1}$ with $m$ and $m+1$ `red' edges are neighbors in $C$ if they differ at a single  edge. If $c_m$ and $c_{m+1}$ are neighbors, it is  clear that $gc_m$ and $gc_{m+1}$ are also neighbors for each $g \in O_h$. Thus, every coloring in the orbit $O_h c_{m}$ is the neighbor of at least one coloring in the orbit $O_h c_{m+1}$ and it is natural to  say that $[c_m]$ and $[c_{m+1}]$ are neighbors in $C/O_h$.

\subsection{A phenomenological energy for molecular distortion}
Each configuration $[c] \in C/O_h$ determines the unique combinatorial structure of a supramolecular conformation. An {\em energy landscape\/} is a map $F: C/O_h \to \mathbb{R}$. 

\begin{figure}
\centering
\includegraphics[width=0.75\textwidth,natwidth=768,natheight=256]{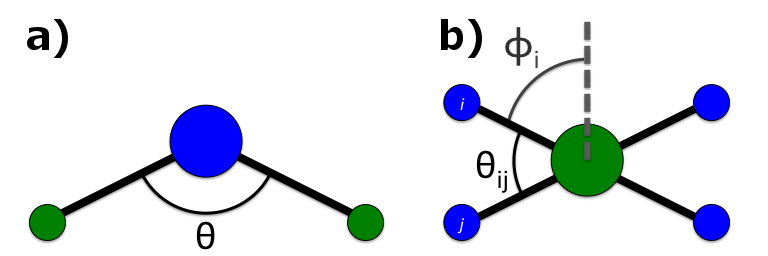}
\caption{Geometry of a) edges, comprising an `elbow' and two arms; and b) 4-valent vertices.}
\label{Fig_model}
\end{figure}

We define a phenomenological energy by penalizing the distortion both of the edges and of the vertices from their preferred geometry.  The edges are modeled as two straight arms, joined at an angle at the `elbow' of the boomerang (Fig.~\ref{Fig_model}a).  
Let $X \in \mathbb{R}^{3|\mathcal{E}| + 3|\mathcal{V}|}$ denote the coordinates of the elbows of the ligands and of the vertices of the embedded polyhedron.  We introduce energy terms which account separately for the distortion from a preferred length for each arm of an edge ($E_{arms}$); distortion from the preferred bend angle at each elbow ($E_{elbow}$); and distortion from square-planar geometry at each vertex ($E_{vertex}$):
\begin{equation}
\label{eq:energy1}
G(X;[c]) =   \sum_{\edge \in \mathcal{E}} \left( E_{arms}(\edge(X)) + E_{elbow}(\edge(X)) \right )  + \sum_{\mathfrak{v} \in \mathcal{V}}E_{vertex}(\mathfrak{v}(X)).
\end{equation}
Here the notation $\edge(X)$ and $\mathfrak{v}(X)$ implies that the energy depends on the coordinates of each edge and vertex respectively. Finally, the energy landscape $F ([c])$ is defined by
\begin{equation}
\label{eq:energy}
F([c]) = \inf_{X} G(X;[c]),
\end{equation}	
where the infimum is taken over all embeddings of $\mathcal{P}$ into $\mathbb{R}^3$. In our numerical experiments, the energy $G$ is minimized through a zero-temperature Monte-Carlo annealing. This annealing does not reach the exact energy minimum, but gives an upper bound on the minimum-energy which we believe to be typically within a few percent of the actual infimum. This fact may be verified for the minimum energy configurations shown in Fig.~\ref{Fig:polyhedra} (d--f).

We now describe the terms in the energy in local coordinates in more detail.  We introduce an energy cost to deviations of the length of the arms from one:
\begin{equation}
\label{eq:arm-energy}	
	E_{arm} = E_{a} ((l_1 - 1.0)^2 + (l_2 - 1.0)^2),
\end{equation}	
where $l_1$ and $l_2$ are the lengths of the two arms.  We choose  $E_{a} \gg E_e,E_p,E_v$ (defined below) to make the arms stiff; this has the effect that in general in the minimum-energy embedding, the arms are of length one and do not contribute signficantly to the total energy.

We also introduce an energy cost to deviations of the angle of the elbow from its preferred opening angle:
\begin{equation}
\label{eq:theta-def}	
E_{elbow} = E_{e} (\theta - \theta_0)^2
\end{equation}
with $\theta$ the actual opening angle and $\theta_0$ the preferred opening angle (which depends on the type $A$ or $B$).  Thus this term is the only term in the energy which depends on the \emph{configuration} as well as on the embedding.  The coefficient $E_{e}$ sets the scale of the energy contribution of the edges.  

At each vertex, we introduce an energy cost for deviations from square-planar geometry (see Fig.~\ref{Fig_model}b):
\begin{equation}
\label{eq:Edef}	
	E_{vertex} = E_v \left [ \sum_{\langle i,j \rangle} (\theta_{ij} - \pi/2)^2 + E_{p} \sum_{i} (\phi_{i} - \pi/2)^2 \right ],
\end{equation}	
with the first sum over pairs of neighboring edges $i$ and $j$ and $\theta_{ij}$ the angle formed by the `arms' of those edges; and the second sum over edges $i$ and $\phi_{i}$ the angle formed between the arm of that edge and `plane perpendicular', defined as the average of the cross-products of the pairs of neighboring edges.  The first sum pushes the edges to all be perpendicular to one another (square), while the second sum pushes the edges to be co-planar.  The overall coefficient $E_v$ sets the scale of the energy contribution of the vertices relative to the edges, while the coefficient $E_{p}$ sets the scale of the energy contribution of the `planar' aspect of the vertices relative to the `square' aspect.

For each polyhedron, there is an ideal angle $\theta_*$ such that there is a zero-energy state when all edges have $\theta_0 = \theta_*$.  We calculate these ideal angles analytically for each of the three polyhedra considered.  Fixing one vertex at the origin, we write explicitly the positions $\mathbf{A}$, $\mathbf{B}$, $\mathbf{C}$, $\mathbf{D}$ of its four neighbors in a regular polyhedron (ordered cyclically so that $\mathbf{B}$ and $\mathbf{D}$ are each closer to $\mathbf{A}$ than is $\mathbf{C}$).  We then solve for $\mathbf{E}$, $\mathbf{F}$, $\mathbf{G}$, $\mathbf{H}$, the positions of the elbows, which satisfy the following conditions:
\begin{gather*}
|\mathbf{E}| = |\mathbf{F}| = |\mathbf{G}| = |\mathbf{H}| \\
\mathbf{G} = -\mathbf{E} \ , \  \mathbf{H} = -\mathbf{F} \\
\mathbf{E} \cdot \mathbf{F} = \mathbf{E} \cdot \mathbf{H} = \mathbf{F} \cdot \mathbf{G} = \mathbf{G} \cdot \mathbf{H} = 0 \\
|\mathbf{A} - \mathbf{E}| = |\mathbf{E}| \ , \  |\mathbf{B} - \mathbf{F}| = |\mathbf{F}| \ , \  |\mathbf{C} - \mathbf{G}| = |\mathbf{G}| \ , \  |\mathbf{D} - \mathbf{H}| = |\mathbf{H}|   \\
\end{gather*}
These relations ensure that the geometry at the vertex is square-planar, and that every arm of an edge is of the same length.  Examining the triangle $\mathbf{OEA}$ we see that the opening angle $\theta$ is then given by
\begin{equation}
\theta_* = \cos^{-1} \left ( 1 - \frac{|\mathbf{A}|^2}{2|\mathbf{E}|^2 } \right )
\label{ideal_angle}
\end{equation}
Thus if the preferred angle of every edge is set to $\theta_*$, the embedding with this geometry at every vertex of the Platonic or Archimedean polyhedron will give a zero-energy configuration; these embeddings are shown in Fig.~\ref{Fig:polyhedra}(d)--(f).  The values derived for the ideal angles of the three polyhedra are:
\begin{align*}
& \theta_{*,\text{octahedron}} & = & \ 90^\circ, \\
& \theta_{*,\text{cuboctahedron}} & = & \ \cos^{-1}\left ( \frac{1-2\sqrt{2}}{4} \right ) & \approx & \ 117.2 ^\circ, \\
& \theta_{*,\text{rhombicuboctahedron}} & = & \ \cos^{-1}\left ( \frac{25-20\sqrt{2}}{33-20\sqrt{2}} \right ) & \approx & \ 134.1^\circ .
\end{align*}
Note that these values for the cuboctahedron and rhombicuboctahedron correct the approximate values of 120$^\circ$ and 135$^\circ$, respectively, given in \cite{Fujita1}.

However, in experiment this value can only be chosen approximately, and the angle $\theta_0$ is different from the ideal value. This causes frustration, as shown in Fig.~\ref{Fig:polyhedra}(g)--(i), and the minimum-energy state balances competing energies.  In these examples, the total energy of the vertices is between 1.8-2.0 times the total energy of the edges.  All of the energy contributions are necessary to observe frustration. This can be seen by counting the degrees of freedom.  For example, if the planarity of the vertices is relaxed, the edge `elbows' can rearrange themselves so that they satisfy their preferred angles at the expense of the square-planar geometry at the vertices.  Only when all four energy terms are in competition do we have an interesting system. It is in this sense that  we consider our energy model to be minimal.

Finally, we define the \emph{average} preferred angle for mixtures of two ligands.  The preferred angle used in \cite{Fujita1} is:
\begin{equation}
\label{eq:mean-angle}
\bar\theta_0(c_A, c_B) = \frac{1}{c_A + c_B} \left( c_A \theta_0(A) + c_B \theta_0(B)	\right), 
\end{equation}
where $c_A$ and $c_B$ denote the concentrations of ligands $A$ and $B$.  We will also use $\bar\theta_0$ with no arguments to mean 
\begin{equation}
\label{eq:mean-angle-half}
\bar\theta_0 \equiv \bar\theta_0(c,c) = \frac{1}{2} \left( \theta_0(A) + \theta_0(B) \right)
\end{equation}
and $\bar\theta_0(m)$ in an isomer $M_nA_{2n-m}B_m$ to mean
\begin{equation}
\label{eq:mean-angle-m}
\bar\theta_0(m) \equiv \bar\theta_0(2n-m,m) = \frac{1}{2n} \left( (2n-m) \theta_0(A) + m \theta_0(B) \right)
\end{equation}

%% file: sampling
\section{Symmetries and Sampling Strategies}
As discussed above, our goals are twofold: 1) to enumerate the configuration space $C/O_h$; and 2) to calculate the energy landscape $F(C/O_h)$.  The computational difficulty of these tasks depends strongly on the size of the polyhedron $\mathcal{P}$.  When $\mathcal{P}$ is the octahedron, we are able to fully enumerate the configuration space  and its graph structure, and compute the entire energy landscape.  When $\mathcal{P}$ is the cuboctahedron, we fully enumerate the configuration space and its graph structure, but we sample the energies of only a fraction of the configuration space to compute a partial energy landscape.  When $\mathcal{P}$ is the rhombicuboctahedron, even the configuration space is intractable.  We enumerate only those configurations $c$ with non-trivial symmetry number $s$, and sample the energy of only a fraction of these configurations.  In what follows, we focus mainly   on the sampling strategies used to obtain partial, but informative, descriptions of the configuration space and energy landscape for the two larger polyhedra.

The octahedron is the most straightforward.  With only $2^{12} = 4096$ colorings to consider, we construct every coloring, and calculate the equivalences under $O_h$ to obtain a complete list of configurations.  There are 144 configurations, of which 94 have non-trivial symmetry groups (see Table~\ref{Table_symmetries}).  We also calculate explicitly the graph structure, identifying the neighbor relations between configurations.  We then calculate the energy of every one of the 144 configurations by solving the minimization problem~(\ref{eq:energy}).

\begin{table}[h]
\centering
\begin{tabular}{| c | c | c | c |}
\hline
\parbox[t]{0.15\textwidth}{\centering symmetry number} & \parbox[t]{0.2\textwidth}{\centering octahedral\\configurations} & \parbox[t]{0.2\textwidth}{\centering cuboctahedral\\configurations} & \parbox[t]{0.3\textwidth}{\centering rhombicuboctahedral\\configurations} \\
\hline
48 & 2 & 2 & 4 \\
24 & 0 & 1 & 4 \\
16 & 2 & 2 & 28 \\
12 & 2 & 6 & 29 \\
8 & 6 & 23 & 644 \\
6 & 4 & 35 & 716 \\
4 & 21 & 211 & 46,991 \\
3 & 1 & 44 & 16,015 \\
2 & 56 & 5,943 & $>$40,648,786 \\
1 & 50 & 346,477 & $\approx 6 \times 10^{12}$ \\
\hline
total & 144 & 352,744 & $\approx 6 \times 10^{12}$ \\
\hline
\end{tabular}
\caption{Enumeration of configurations by symmetry number for each polyhedron.}
\label{Table_symmetries}
\end{table}

We build the graph $C/O_h$ for the cuboctahedron  by induction on the number of `red' edges, denoted $m$.  There is a single configuration which has $m=0$, that is, all edges are `blue'; this provides the base case.  As our induction step, we use the subset $(C/O_h)_m$ of configurations with $m$ `red' edges to construct the set $(C/O_h)_{m+1}$ as follows.  For each element $[c] \in (C/O_h)_m$, we consider a representative coloring $c \in [c]$. This coloring has neighbors $\{c'\} \subset C_{m+1}$, obtained from $c$ by changing the color of a single edge from `blue' to `red'. We compute the orbit $O_hc'$ for each neighbor $c'$,  and compare each element of $O_hc'$ against the configurations already seen in $(C/O_h)_{m+1}$.  If the configuration has already been seen in $(C/O_h)_{m+1}$, we make note that $[c]$ and $[c']$ are neighbors. If the configuration has not been seen before, we update our list of representatives in $(C/O_h)_{m+1}$ to include $[c']$ and make note of the neighbor relationship.  This process ends when we find the single configuration with $m=2n=24$.  Thus we do not enumerate every possible coloring, but we nonetheless identify all configurations and  compute the connectivity of the graph $C/O_h$.  We find a total of 352,744 configurations (see Table~\ref{Table_symmetries}).

The computation of the energy for each configuration $[c]$ involves a numerical solution of the minimization problem~(\ref{eq:energy}). Rather than pursue this computation for each of the 352,744 configurations, we sample a fraction of the configurations and compute a partial energy landscape.  We use a sampling strategy which is designed to identify the full range of energies at each value of $m$, that is, to find the lowest and highest energy configurations at each $m$.  
The strategy exploits two heuristic observations about the energy landscape of the octahedron: 
\begin{enumerate}
\item For a given $m$, the highest and lowest energy configurations tend to have nontrivial symmetry number $s$; 
\item High energy configurations tend to have high energy neighbors in the graph.   Likewise, low energy configurations have low energy neighbors.  
\end{enumerate}

Our sampling strategy consists of two steps.  First, we calculate the energy for all configurations with `extreme' values of $m$, here defined as $m\leq4$ or $m\geq20$ (630 configurations total), and for all configurations with symmetry number $s\geq 3$ (324 configurations total, although there is overlap with the 630 extreme-$m$ configurations).  Finally, we calculate the energies of about $2\%$ (but at least $100$) of the remaining configurations for each value of $m$ such that $5 \leq m \leq 19$, chosen as follows.  About $\frac{1}{3}$ to $\frac{1}{2}$ of these configurations are identified as the neighbors of the highest and lowest energy configurations known at $m-1$ and $m+1$.  The other roughly half of these configurations are sampled uniformly at random from the remaining configurations.  In order to maximize the information we have about neighbors of high and low energy configurations, we carry out this calculation sequentially for different values of $m$, starting with $m=5$ and $m=19$ and working our way to less extreme values, ending with $m = n = 12$.

For the rhombicuboctahedron, the combinatorial explosion in the number of configurations makes it intractable to even enumerate the approximately six trillion configurations.  We focus instead on configurations with non-trivial symmetries,  motivated by the heuristics above. 

A configuration with a non-trival symmetry number $s \geq 2$ is fixed by at least one non-identity element $g \in O_h$.  We generate such `symmetric' configurations by first enumerating the set of colorings fixed by $g$, $C^g$, for each $g \in O_h$, $g \neq e$ as follows.  We consider the orbits of the \emph{edges} of the polyhedron under the subgroup $S_g = \langle g \rangle \subset O_h$ generated by $g$.  In order for a coloring $c$ to be fixed by $g$, i.e.~$gc = c$, the edges in each  orbit $S_g\edge$, $\edge \in \mathcal{E}$ must be of the same color.  Thus, $C^g$ is constructed as the set of 2-colorings of edge orbits, $S_g \edge$. We classify the forty-seven non-trivial elements of $O_h$  into the following five subsets:


\begin{enumerate}
\item[(i)] There are 20 elements of $O_h$ with order $\geq 4$.  Each of these elements has a power $g^a$ which is of order 2 or 3.  Since any coloring fixed by $g$ is also fixed by $g^a$, $C^{g} \subset C^{g^a}$, and we need only enumerate $C^g$ for $g$ of order 2 or 3 to get all configurations.

\item[(ii)]  There are 8 elements of $O_h$ with order 3; while these fix different colorings, they generate the same equivalence classes in $C/O_h$.  Thus we need enumerate $C^g$ for only one of these 8 elements. 

\item[(iii)] There are 10 elements of $O_h$ with order 2 which have the property that $|S_g\edge| = 2$ for all $\edge\in \mathcal{E}$.  Of these only three produce different configurations, while the remaining seven fix colorings which are equivalent to colorings already identified. 

\item[(iv)] There are 6 elements of $O_h$ with order 2 which have four orbits for which $|S_g\edge| = 1$, that is, these elements map four edges onto themselves.  As with the elements of order 3, these produce the same configurations, and so we enumerate $C^g$ for only one.

\item[(v)] There are 3 elements of $O_h$ with order 2 which have eight orbits for which $|S_g\edge| = 1$, that is, these elements map eight edges onto themselves.  These elements represent reflections across the plane bisecting a ring of square faces. 
Again, we enumerate $C^g$ for only one of them.
\end{enumerate}

Having enumerated $C^g$ for several elements $g$, we obtain a list of colorings which contains at least one representative of $[c]$ for each $[c] \in C/O_h$ such that $s \geq 2$.  We then compare these colorings and their equivalences to enumerate all $[c]$ with $s \geq 3$ (64,431 configurations total; see Table~\ref{Table_symmetries}).  There are tens of millions of configurations with symmetry $s = 2$; we have not enumerated all of these, but have enumerated those with extreme values of $m$ ($m\leq8$ and $m\geq40$), along with those at values of $m$ relatively prime to 48 which have no higher-symmetry configurations ($m=11,13,17,19,23,25,29,31,35,37$).

We then calculate energies for configurations sampled as follows.  We calculate the energy for all configurations with symmetry $s\geq6$ (1425 configurations total); about 2.5\% (but at least 40 at each $m$) randomly selected configurations with $s=4$; about 2.5\% (but at least 40) randomly selected configurations with $s=3$; and for $m\leq8$, $m\geq40$, or $m$ relatively prime to 48, about 0.02\% (but at least 200) configurations with $s=2$.\footnote{In Fig.~\ref{Fig_variations}a, we calculate the energy for configurations with $s=3$ only for $m\leq8$ and $m\geq40$, and for a larger proportion of configurations with $s=2$ for $m\leq8$, $m\geq40$, $m=11,13,17,31,35,37$.}  This gives us a total of roughly 10,000 energies calculated for each of the plots in the next section -- a small fraction of the 6 trillion configurations, but a sampling which we believe captures the most interesting aspects of the energy landscape.

%% file: results
\section{Results}
\label{sec:results}
In most of the examples below, we set the energy parameters $E_a = 10.0$; $E_p = 1.0$; and $E_e = E_v$ with $1.0 < E_e < 3.8$, with the precise value typically chosen to normalize the energy of the $m=0$ configuration to 1.0.  In Section~\ref{Section:energy_parameters}, we describe results with $E_e \neq E_v$.

\subsection{Energy landscape when $\bar\theta_0 = \theta_*$}
\begin{figure}
\centering
\includegraphics[width=\textwidth,natwidth=1080,natheight=1080]{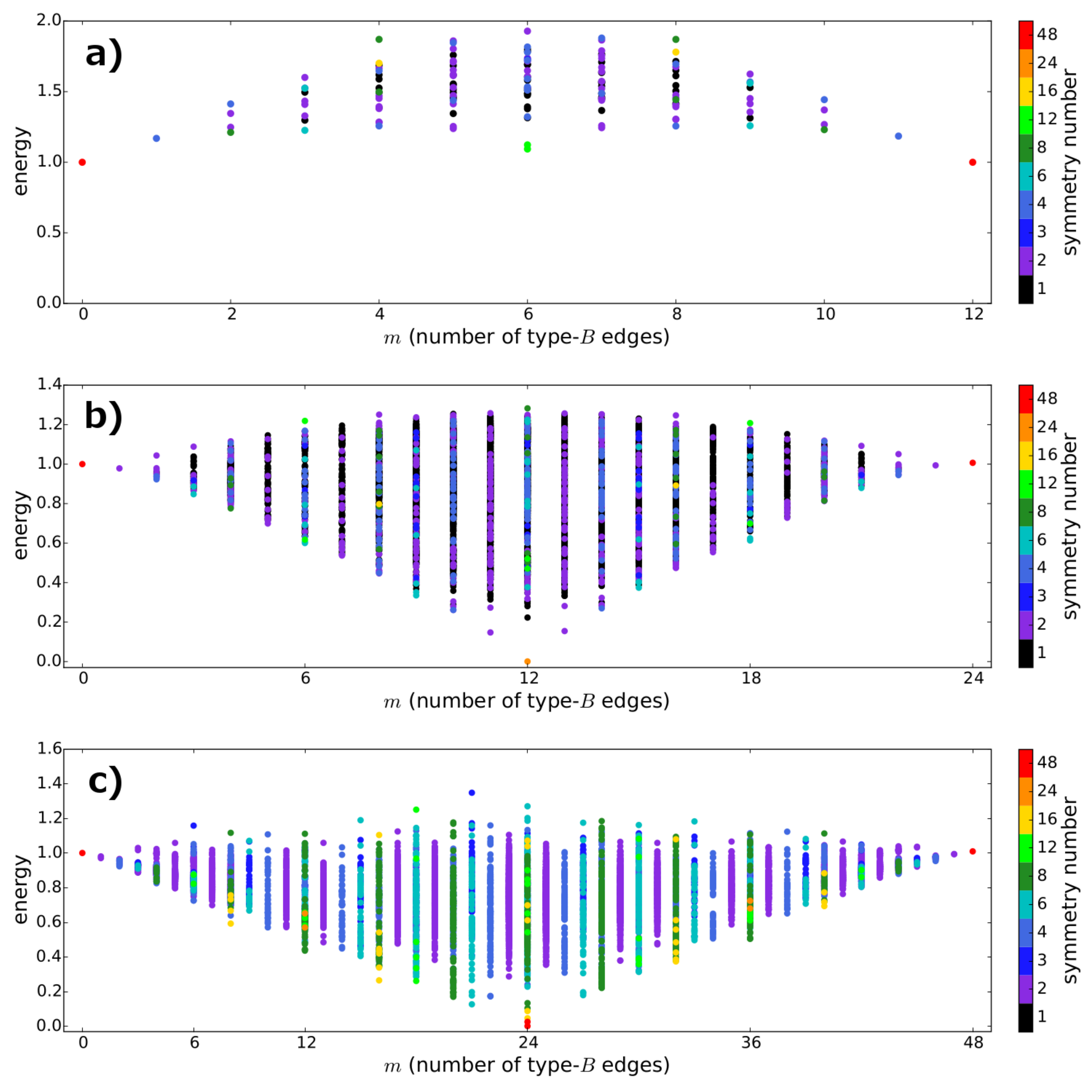}
\caption{{\bf Energy landscapes and partial energy landscapes\/} for a) the octahedron; b) the cuboctahedron; and c) the rhombicuboctahedron.  For each polyhedron, $\theta_0(A)$ and $\theta_0(B)$ are chosen so that ${\theta_0(A) < \theta_* < \theta_0(B)}$, and ${\bar\theta_0 = \theta_*}$: a) ${\theta_0(A) = 81^\circ}$, ${\theta_0(B) = 99^\circ}$ (${\theta_* = 90^\circ}$); b) ${\theta_0(A) \approx105.5^\circ}$, ${\theta_0(B) \approx 128.9^\circ}$ (${\theta_* \approx 117.2^\circ}$); c) ${\theta_0(A) \approx 120.7^\circ}$, ${\theta_0(B) \approx 147.6^\circ}$ (${\theta_* \approx 134.1^\circ}$).  The energies are normalized so that the configuration with $m=0$ has energy 1.0.  For (a), the entire energy landscape is shown; for the partial landscapes in (b) and (c), the choice of sampled energies is described in the text.  Color indicates the symmetry number of the configuration.}
\label{Fig_spectra}
\end{figure}

In Fig.~\ref{Fig_spectra}, we present sample energy landscapes for the octahedron, the cuboctahedron, and the rhombicuboctahedron, when the average preferred angle is the same as the ideal angle, that is, $\theta_* = \bar\theta_0 = \frac{1}{2}\left ( \theta_0(A)+ \theta_0(B) \right )$. (This choice of parameters was used as a heuristic design principle in~\cite{Fujita1}; as discussed below, the broad features of this energy landscape are largely insensitive to these choices of $\theta_0(A)$ and $\theta_0(B)$.)
We observe that when $n$ and $m$ are both held fixed, there is nonetheless a strong variation in the energy of all isomers with chemical formula $M_{n}A_{2n-m}B_{m}$.  The overall maximum energy is found at $m=n$, as is the minimum energy for both the cuboctahedron and the rhombicuboctahedron; this minimum energy appears to be zero, within the error of our minimization.  For the octahedron, the minimum energy is 1.0 at $m=0$ or $m=2n$, but there is a low-energy configuration at $m=n$ with only slightly higher energy.

\subsection{Extremal energy configurations have high symmetry}
This low-energy configuration at $m=n$ is also the isomer with the largest symmetry number.  While this may seem unsurprising, it must be recognized that there are also high-energy configurations with high symmetry number!

More generally, in all of the energy landscapes shown in Fig.~\ref{Fig_spectra}, for any given $m$, the configurations with non-trivial symmetry number tend to span or nearly span the range of energies.  For the octahedron, for which the energy landscape is completely calculated, the extremal energy configurations at each $m$ have symmetry number $s\geq2$; indeed, for $m\neq5,7$, the lowest-energy configuration has $s\geq4$.  For the cuboctahedron, we have calculated only a partial energy landscape, with our sampling biased toward high-symmetry configurations, but nonetheless we find similar results.

\begin{figure}
\centering
\includegraphics[width=\textwidth,natwidth=788,natheight=525]{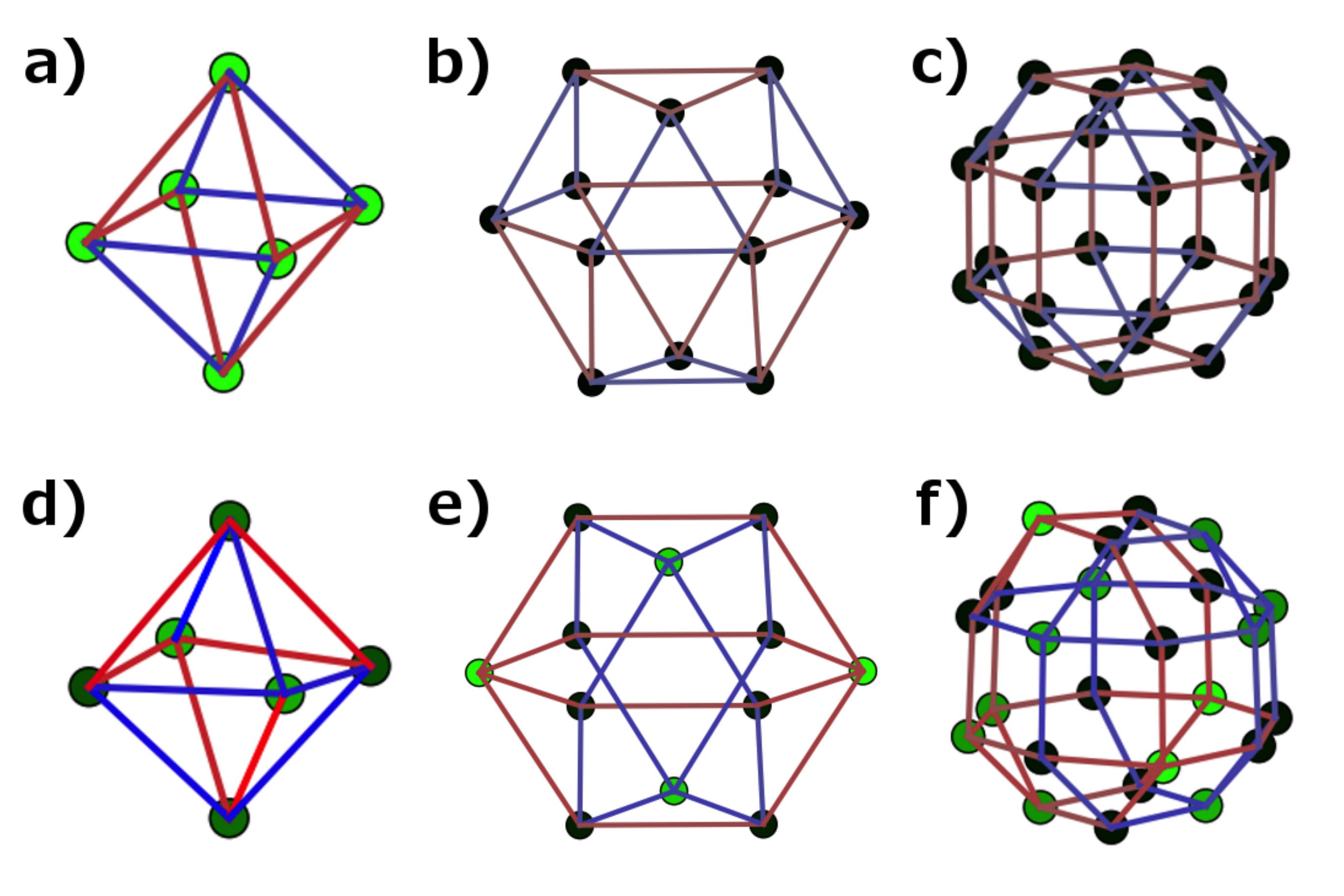}
\caption{{\bf Extremal energy configurations for $m=n$\/}.  For clarity, the edges are drawn as straight rather than showing the full embedded cage structure.  The preferred angles are as in Fig.~\ref{Fig_spectra}; blue edges are of type $A$ with $\theta_0(A) < \theta_*$, while red edges are of type $B$ with $\theta_0(B) > \theta_*$.  Brightness of the color indicates the energy contribution from each vertex and each edge; the color scale is different in each panel.  Note the deformation from a regular polyhedron; in particular, blue edges are shorter than red edges.  a-c) Minimum-energy configurations for a) the octahedron; b) the cuboctahedron; and c) the rhombicuboctahedron.  d-f) Maximum-energy configurations.  Note the high symmetry numbers: a) $s=12$; b) $s=24$; c) $s=48$; d) $s=2$; e) $s=6$; f) $s=6$.}
\label{Fig_extremal_energies}
\end{figure}

Some insight into why symmetry plays such an important role may be gained by considering the structure of these extremal-energy configurations, shown in Fig.~\ref{Fig_extremal_energies} for $m=n$ for each polyhedron.  Consider first the low-energy isomers: in all three polyhedra, at every vertex meet two edges of type $A$ and two edges of type $B$ in a ``\emph{cis}'' arrangement such that the edges of type $A$ are adjacent to one another, and opposite the edges of type $B$.  This arrangement allows the frustration to be resolved by canting the plane of the ion (the resolution is not complete for the octahedron, since canting fully would require stretching the stiff `arms').  Heuristically, then, for other values of $m$, we expect that configurations with many vertices with this energetically favorable configurational structure will have low energies, and the similarity of vertices tends to yield high symmetry number.  In the case of the high-energy isomers, those for the cuboctahedron and the rhombicuboctahedron show very similar configurational structure: they are composed of 6- or 8-element closed rings which are either entirely of type $A$ or entirely of type $B$, so that at every vertex meet either four edges of the same type, or two edges of each type in a ``\emph{trans}'' arrangement such that edges of the same type lie opposite one another.  The largest energy contributions are from vertices at which four edges of type $B$ meet, followed by those at which four vertices of type $A$ meet.  Again, configurations with many vertices with these energetically unfavorable structures will have high energies.

\subsection{Low and high energy configurations are neighbors in the graph structure}
\begin{figure}
\centering
\includegraphics[width=\textwidth,natwidth=1080,natheight=360]{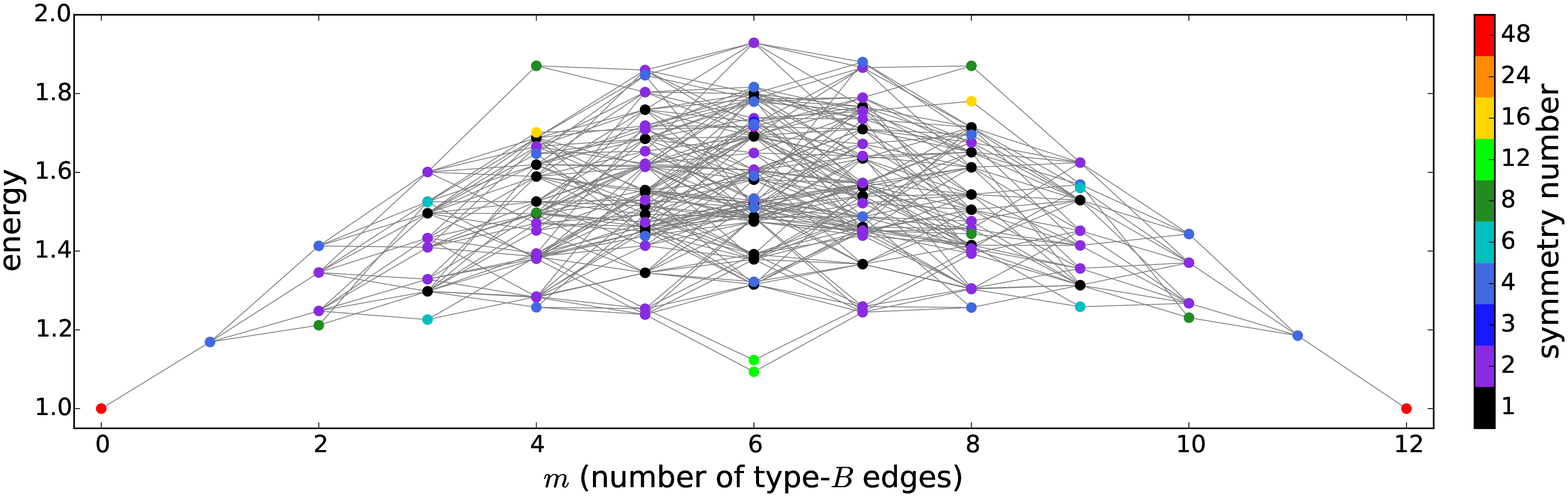}
\caption{Energy landscape of the octahedron shown with the graph structure of the configuration space.  (The range of the $y$-axis has been changed from Fig.~\ref{Fig_spectra}a for clarity).  Note that low energy configurations tend to be connected to other low energy configurations, and likewise high energy configurations to high energy configurations.}
\label{Fig_config_graph}
\end{figure}

For the octahedron, we obtain a complete description of  the configuration space as a graph, as shown in Fig.~\ref{Fig_config_graph}.  We observe that low (respectively high) energy configurations are usually neighbors of other low (resp. high) energy configurations. This feature of the energy landscape may be explained heuristically as follows. Starting from a low (resp. high) energy configuration with several favorable (resp. unfavorable) vertex arrangements as described above, changing the type of a random edge will leave many of these vertex arrangements unchanged, resulting in a modest change in energy and giving another low (resp. high) energy configuration. 

\subsection{Energy landscape when $\bar\theta_0 \neq \theta_*$}
\begin{figure}
\centering
\includegraphics[width=\textwidth,natwidth=1080,natheight=1080]{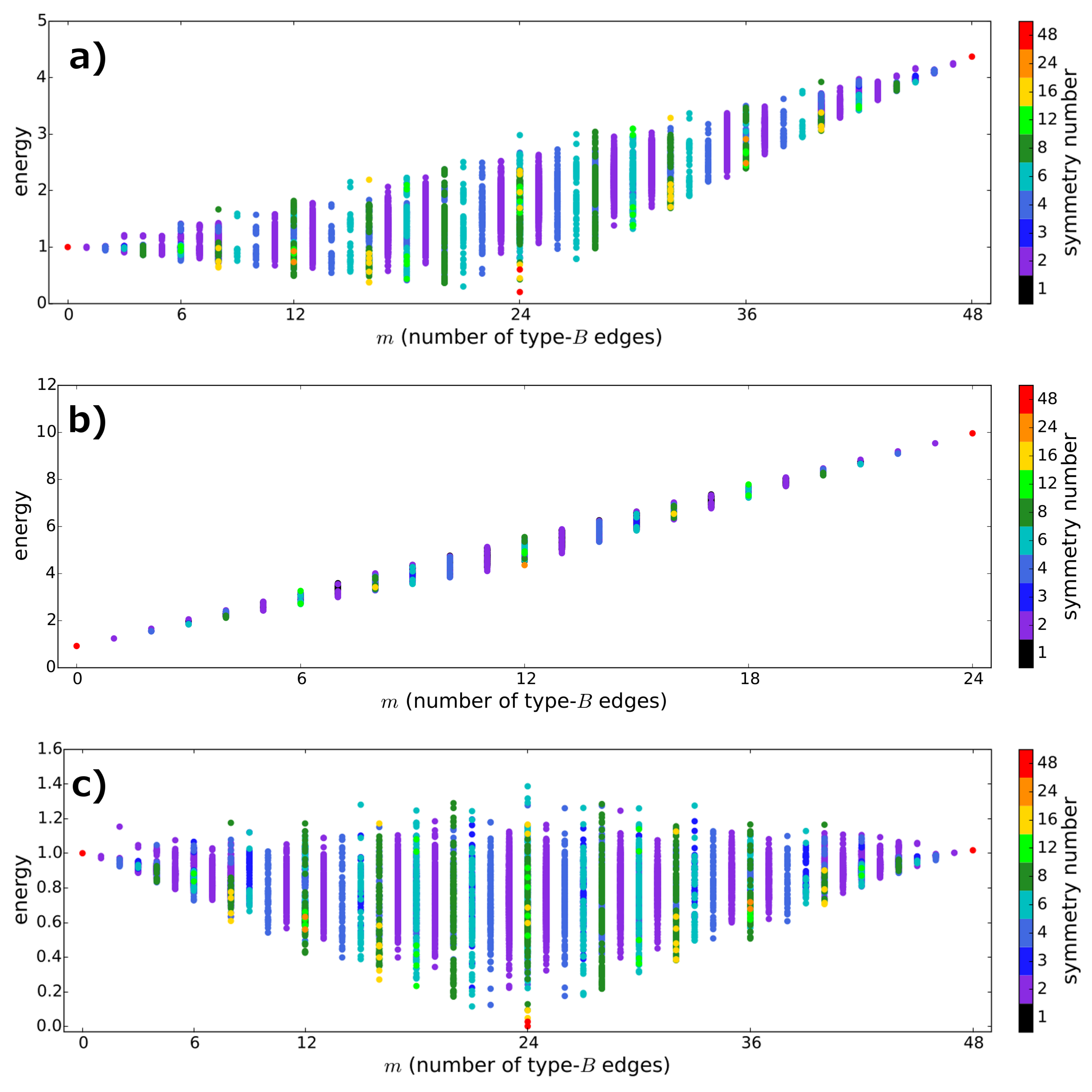}
\caption{{\bf Variation of the energy landscapes with model parameters\/}.  a-b) {\bf Change of preferred angles\/}.  Partial energy spectra for a) the rhombicuboctahedron and b) the cuboctahedron with preferred angles $\theta_0(A) = 127^\circ$, $\theta_0(B) = 149^\circ$; these angles are chosen to match those of the two organic ligands reported in \cite{Fujita1}.  For the rhombicuboctahedron (a), $\theta_0(A) < \theta_* < \theta_0(B)$ still holds, but $\bar\theta_0(m) = \theta_*$ for $m \approx 16$; note that the lowest energy configuration is still at $m=n=24$.  For the cuboctahedron (b), $\theta_* < \theta_0(A) < \theta_0(B)$; note that the energy increases rapidly with added type-$B$ edges.  c) {\bf Change of relative energy scales\/}. Partial energy spectrum for the rhombicuboctahedron with the same preferred angles as in Fig.~\ref{Fig_spectra}c, but with $E_e = 2E_v$.  Note that the shape of the energy landscape is almost unchanged.  For (a) and (c), the overall energy scale is chosen so that the configuration with $m=0$ has energy 1.0.}
\label{Fig_variations}
\end{figure}

We find that the shape  of the energy landscape is robust for variations of $\theta_0(A)$ and $\theta_0(B)$.  When $\theta_0(A) < \theta_* < \theta_0(B)$, but ${\theta_* \neq \bar\theta_0 = \frac{1}{2}(\theta_0(A) + \theta_0(B))}$, the energy spectrum becomes tilted as shown in Fig.~\ref{Fig_variations}a, with $$\frac{E(m=2n)}{E(m=0)} \approx \frac{|\theta_0(B) - \theta_*|^2}{|\theta_0(A) - \theta_*|^2}.$$  Importantly, the lowest energy is still found at even mixing, $m=n$, even though $\theta_0(m=n) \neq \theta_*$; the energy is nonzero, but still small compared to the energy at $m=0$ and $m=2n$.  We conclude that even if $\bar\theta_0 \neq \theta_*$, the energetically favorable arrangements of type A and type B edges described above allow for much of the frustration to be relieved.

The lowest energy configuration instead occurs for $m=0$ when ${\theta_* < \theta_0(A) < \theta_0(B)}$, and the energy increases roughly linearly with $m$, as shown for the cuboctahedron in Fig.~\ref{Fig_variations}b.  Isomers with a fixed $m$ still exhibit a range of energies, but the variation with $m$ is similar to the $m$-dependent energy contribution.

\subsection{Independence of energy scales}
\label{Section:energy_parameters}
There is little variation in the shape of the energy spectrum as the parameters $E_e$, $E_v$, and $E_p$ are varied (we fix $E_a = 10.0 > E_e, E_v, E_p$). Fig.~\ref{Fig_variations}c shows the energy spectrum for a rhombicuboctahedron with the same preferred angles as in Fig.~\ref{Fig_spectra}c, such that $\bar\theta_0 = \theta_*$, but with $E_e = 2E_v$ ($E_p$ is kept at 1.0).  This is nearly indistinguishable in overall shape from Fig.~\ref{Fig_spectra}c, with $E_e = E_v$; there is a slight increase in the maximal energies reached but it is subtle.  Similar results were obtained for the cuboctahedron with $E_e=\frac{1}{2}E_v$; again the energy landscape has much the same shape, with in this case a slight decrease in the maximal energies (data not shown).

We also observed that the ratio between the energy of a rhombicuboctahedron with only one ligand type, $M_{24}A_{48}$ and a cuboctahedron with the same ligand type and parameters, $M_{12}A_{24}$, is nearly constant over a much larger range of energy ratios, varying both $E_e/E_v$, and $E_p$ (data not shown).  Thus the ratios of the energy parameters in this system seem to be relatively unimportant: the competition between the edge energies and vertex energies results in an embedding which balances those energies in such a way that the shape of the energy spectrum is roughly independent of the ratios, and this shape is almost entirely determined by the preferred angles.  The overall energy scale, of course, will be important in any system in which temperature plays a role.

\subsection{Crossover of landscapes for different polyhedra} 
\begin{figure}
\centering
\includegraphics[width=\textwidth,natwidth=1080,natheight=360]{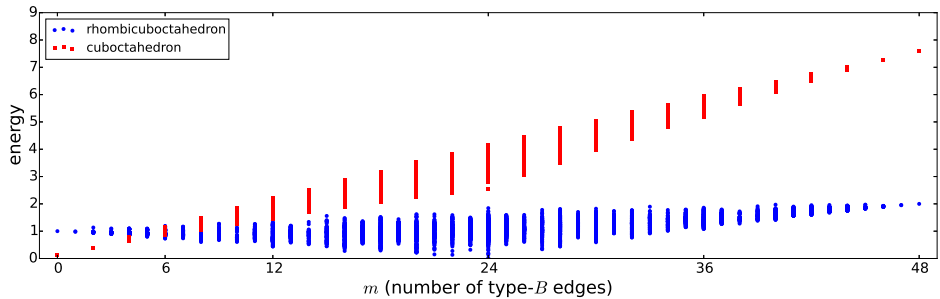}
\caption{{\bf Crossover of cuboctahedral and rhombicuboctahedral landscapes.\/}
Partial energy landscapes of the cuboctahedron and the rhombicuboctahedron, both with preferred angles ${\theta_0(A) = 121.5^\circ}$, ${\theta_0(B) = 151.8^\circ}$; note ${\theta_{*,cuboct.} < \theta_0(A) < \theta_{*,rhombicuboct.} < \theta_0(B)}$.  Blue circles give the landscape of the rhombicuboctahedron, and red squares that of the cuboctahedron.  The landscape of the cuboctahedron is modified by doubling both $m$ and the energy, giving in effect the total $m$ and total energy of two identical cuboctahedra; this allows us to preserve the total number of ions and of ligands when we compare the rhombicuboctahedron with the smaller cuboctahedron.  The energies are normalized so that the rhombicuboctahedron with $m=0$ has energy 1.0.}
\label{Fig_cuboct_rco_comparison}
\end{figure}

We compare two different polyhedra to one another as shown in Fig.~\ref{Fig_cuboct_rco_comparison}.  For the cuboctahedron, we scale both $m$ (the number of `red' edges) and the energy by two in order to compare energy per ligand.  The preferred angles are $\theta_0(A) \approx121.5^\circ$ and $\theta_0(B)\approx151.8^\circ$, so that for the cuboctahedron, $\theta_{*,cuboct.} < \theta_0(A) < \theta_0(B)$, while for the rhombicuboctahedron, $\bar\theta_0(m) = \theta_{*,rhombicuboct.}$ for $m\approx20$.  Note that there is a crossover; for $m\lesssim4$, the cuboctahedron is energetically favored, while for $m\gtrsim8$, the rhombicuboctahedron is energetically favored.  This is suggestive that as the concentrations of type $A$ and type $B$ ligands are varied, a transition may be observed from the cuboctahedron to the rhombicuboctahedron, as in \cite{Fujita1}, although in our model, mixing of the two polyhedra is suggested for $m\approx6$, which was not observed in \cite{Fujita1}.

%% file: conclusions
\section{Discussion}
\label{sec:conclusions}

We have studied in some detail the energy landscape  for a minimal model of supramolecular polyhedra formed by mixtures of two types of ligands. We have found a strong variation with configuration in the energy of different isomers, configurations which have the same number of type-$A$ and of type-$B$ ligands.  When the preferred angles encompass the ideal angle for the polyhedron, that is, $\theta_0(A) < \theta_* < \theta_0(B)$, the lowest energies are achieved by high-symmetry configurations in which edges meet at {\em cis\/} configurations at a vertex, with a very low energy configuration at $m=n$. While our work does not fully explain the emergence seen in~\cite{Fujita1}, for some choices of parameters we see a shift in minimum energy states from cuboctahedron to rhombicuboctahedron. We also note a striking analogy to another experiment in Fujita's lab, in which ligands with different lengths but similar angles were mixed \cite{Fujita3}. For a sufficiently large ratio between the lengths, $r\approx 2$, they observed a mixed yet sorted structure of the form $M_{12}A_{12}B_{12}$ with a well-defined structure in which ligands of type $A$ also took on a \emph{cis} configuration opposite ligands of type $B$.  Our work shows that similarly mixed yet sorted structures are favored when ligands with different angles are mixed, although the energy differences are small enough that the sorting is not as complete as in \cite{Fujita3}.

Our model can easily be extended to a system with temperature by introducing two additional terms to obtain the free energy.  The first term is of the form $k_BT/s$, where $s$ is the symmetry number of the configuration; at high enough temperature, this term will change the relative free energies of configurations at a fixed value of $m$.  The second term is of the form $k_BT (1-p)^{2n-m} p^m$, where $1-p$ and $p$ are the bulk proportions of ligands of type $A$ and $B$, respectively; this term depends only on $m$ and not on the configuration, but at high enough temperature will favor configurations with compositions near $M_nA_{2(1-p)n}B_{2pn}$.  These temperature-dependent entropic terms must be significant in the experimental results of \cite{Fujita1}, which report polyhedra with a range of values of $m$, indicating that states other than the ground state are populated.

Further improvements to the model would include the introduction of chemically relevant values for the energy scales $E_a, E_e, E_v$ and $E_p$, which would provide a more quantitative comparison to the experimental results.  Also desirable would be a description of the kinetics; we have so far considered only the energy landscape, but the question of how the edges of different types dynamically arrange themselves is crucial to establishing whether the system can reach its equilibrium distribution and on what timescale, or whether non-equilibrium states become trapped.


These results provide insight into the importance of configuration and symmetry in the determination of the free energies of self-assembling systems. Our work also illustrates the importance of exploiting symmetry to reduce the size of otherwise combinatorially intractable problems.